\documentclass[conference]{IEEEtran}
\IEEEoverridecommandlockouts
\usepackage{cite}
\usepackage{amsmath,amssymb,amsfonts}
\usepackage{graphicx, color, float}
\usepackage{subcaption}
\usepackage[export]{adjustbox}
\usepackage{textcomp}
\usepackage{url} 
\usepackage[table]{xcolor} 
\usepackage{paralist}
\usepackage{enumitem}
\usepackage{multirow}
\usepackage{hyperref}
\usepackage{algorithmic}
\usepackage{framed}
\usepackage[linesnumbered,ruled,vlined]{algorithm2e}

\hypersetup{
    colorlinks=false,
    pdfborder={0 0 0},  
    hidelinks           
}

\def\BibTeX{{\rm B\kern-.05em{\sc i\kern-.025em b}\kern-.08em 
    T\kern-.1667em\lower.7ex\hbox{E}\kern-.125emX}}

\newcommand{\smallurl}[1]{\footnotesize\url{#1}}

\definecolor{baselinecolor}{gray}{.9}

\begin{document}

\title{MedChat: A Multi-Agent Framework for Multimodal Diagnosis with Large Language Models}

\author{
\IEEEauthorblockN{
Philip R. Liu\textsuperscript{1}, 
Sparsh Bansal\textsuperscript{1}, 
Jimmy Dinh\textsuperscript{1}, 
Aditya Pawar\textsuperscript{1}, 
Ramani Satishkumar\textsuperscript{1}, 
Shail Desai\textsuperscript{1}, \\
Neeraj Gupta\textsuperscript{1},
Xin Wang\textsuperscript{2}, 
Shu Hu\textsuperscript{1*}\thanks{\textsuperscript{*}Corresponding author}
}
\IEEEauthorblockA{
\textsuperscript{1}Purdue University, West Lafayette, IN, USA \\
\texttt{\{liu3688, bansa125, dinh16, aspawar, rsatishk, shdesai, gupt1031, hu968\}@purdue.edu}
}
\IEEEauthorblockA{
\textsuperscript{2}University at Albany, State University of New York, Albany, NY, USA \\
\texttt{xwang56@albany.edu}
}
}

\maketitle
    \thispagestyle{plain}
\pagestyle{plain}

\begin{abstract}
The integration of deep learning-based glaucoma detection with large language models (LLMs) presents an automated strategy to mitigate ophthalmologist shortages and improve clinical reporting efficiency. However, applying general LLMs to medical imaging remains challenging due to hallucinations, limited interpretability, and insufficient domain-specific medical knowledge, which can potentially reduce clinical accuracy. Although recent approaches combining imaging models with LLM reasoning have improved reporting, they typically rely on a single generalist agent, restricting their capacity to emulate the diverse and complex reasoning found in multidisciplinary medical teams. To address these limitations, we propose MedChat, a multi-agent diagnostic framework and platform that combines specialized vision models with multiple role-specific LLM agents, all coordinated by a director agent. This design enhances reliability, reduces hallucination risk, and enables interactive diagnostic reporting through an interface tailored for clinical review and educational use. Code available at \url{https://github.com/Purdue-M2/MedChat}.

\end{abstract}


\section{Introduction}

Glaucoma is the leading cause of irreversible blindness worldwide, demanding early detection and scalable diagnosis strategies \cite{Parihar2016-qw}. However, in practice, screening for glaucoma is hindered by the limited availability of ophthalmologists, especially in resource-limited settings \cite{gan2025ophthusgpt}. This gap has driven extensive research into machine learning-based computer-aided diagnosis (CAD) systems for automating glaucoma detection and monitoring. At the same time, foundation models \cite{hu2025improving} and LLMs have rapidly gained prominence in clinical artificial intelligence (AI) workflows, from answering medical questions to drafting patient reports. Combining the pattern-recognition accuracy of deep learning with the contextual knowledge and communication skills of LLMs offers promising potential for automating diagnostic workflows, such as generating a clinical report from an image in a way that reflects how a clinician would interpret and explain it \cite{Wang2024}. 

However, incorporating LLMs into medical imaging \cite{bansal2025robust,huang2025diffusion,yang2024llm} tasks remains challenging. A key concern is hallucination. LLMs may produce confident-sounding statements that are factually incorrect or not grounded in visual evidence \cite{Kim2025.02.28.25323115}. For instance, an LLM might erroneously describe a symptom or measurement that the image does not support. Additionally, the use of LLMs on imaging data can lack interpretability, making it hard for clinicians to trust the output. Recent studies highlight that current LLMs, even powerful ones like GPT-4, lack specialized medical expertise and can yield unreliable advice when used as diagnostic assistants \cite{Zhao_2024}. Researchers have further emphasized that these models often fail to engage with domain-specific detail, limiting their diagnostic reliability \cite{Zhao_2024}.

Researchers have begun to address these issues by combining medical image analysis \cite{huang2024robustly,tsai2024uu2,tsai2024uu,zheng2024contextual,yang2024explainable,lin2024robust,wang2024challenge,sun2024repmedgraf,zhu2024cgd,hu2024umednerf,wang2024neural,hu2023attention} models with LLM-driven reasoning. Recent systems like ChatCAD and OphthUS-GPT have begun to integrate LLMs into CAD workflows, using them to refine visual outputs into natural language reports \cite{Wang2024, gan2025ophthusgpt}. However, these frameworks typically rely on a single generalist agent, which may struggle to emulate the nuanced, role-specific decision-making of multidisciplinary teams. Relying on a single agent increases the risk of errors, omissions, and shallow reasoning across specialized domains. Therefore, simply pairing a vision model with a generic LLM may yield fluent reports, but with uncertain clinical accuracy and consistency.

In this work, we propose MedChat, a multi-agent diagnostic framework that builds on recent advances in CAD and LLM integration by emulating the collaborative reasoning of real medical teams. Unlike prior single-agent approaches, MedChat assigns distinct diagnostic roles to multiple LLM agents and coordinates their outputs through a director agent. This design encourages diverse clinical perspectives, mitigates hallucination risk, and produces more comprehensive and coherent diagnostic reports. While we demonstrate MedChat for glaucoma diagnosis, the framework generalizes to other imaging-based tasks and supports both medical evaluation and educational use.

To support real-world usability, we also present a companion platform that enables PDF report downloads and question-and-answer interaction with the diagnostic output. By facilitating both structured reporting and dynamic follow-up, MedChat makes AI-generated diagnostics more transparent, verifiable, and accessible for clinical review, patient communication, and medical training.

In summary, we make two main contributions: 
\begin{enumerate}
    \item We propose a novel multi-agent framework that distributes diagnostic reasoning across role-specific LLM agents, coordinated by a director agent to produce clinically grounded reports. 
    \item We develop the MedChat platform to demonstrate our approach end-to-end on fundus images. To our knowledge, it is the first ophthalmic diagnostic system to combine CAD and LLMs in a structured multi-agent setting, extending prior work such as ChatCAD, Med-MLLM, and OphthUS-GPT through a collaborative reasoning paradigm.  
\end{enumerate}

\section{Related Work}
\subsection{Glaucoma Detection}
Glaucoma detection has long been an area of active research. The use of neural networks in its detection dates back to the 1990s \cite{kelman1991neural}. In fundus photography \cite{jackman1886photographing}, detailed images of the retina, optic disc, and blood vessels are captured, and this procedure uses a low-power microscope and an attached ``fundus camera" \cite{zeissclarus500}. In contrast, OCT \cite{huang1991optical} scans use light waves, typically infrared, to scan the eye. They create detailed 3D images that can reveal layers of the retina. These images enable the detection of glaucoma through features such as retinal layer thickness \cite{wollstein2014oct}, vertical cup-to-disc ratio \cite{phene2019deep}, and visual field patterns \cite{brusini2007staging}. These biomarkers have been widely studied and incorporated into both classical and deep learning models for automated glaucoma detection. Modern encoder-decoder architectures have recently been employed for glaucoma detection tasks \cite{Hemelings2023Generalizable, Sharma2025Hybrid}. 

\subsection{Large Language Models in Medical Image Analysis}
Growing confidence in LLMs has spurred a surge of research into critical tasks such as medical report generation. Early efforts concentrated on tight two-stream fusion between images and text. MDLM \cite{yao2024mdlm} pairs a CNN image encoder with a Bi-LSTM-attention text encoder and merges them through a gated fusion layer, enabling joint disease classification, lesion localization, and free-text description from heterogeneous CT, MRI, and ultrasound studies. Their results show consistent gains over single-modal baselines and highlight the value of cross-modal representation learning for end-to-end computer-aided diagnosis. 

More recent systems incorporate LLMs into reasoning loops atop conventional vision backends. For instance, ChatCAD \cite{Wang2024} transforms the outputs of multiple diagnostic models into natural language and uses an LLM to refine the resulting reports and support interactive follow-up questions. ViGPT2 \cite{raminedi2024vigpt2} aims to achieve similar goals through an encoder-decoder architecture that pairs a lightweight Vision Transformer backbone with GPT-2 and retrieval-based augmentation, reducing hallucinations and enhancing semantic faithfulness in IU-Xray reporting.

\section{Method}

\begin{figure*}[t]
    \centering
    \includegraphics[width=1\linewidth]{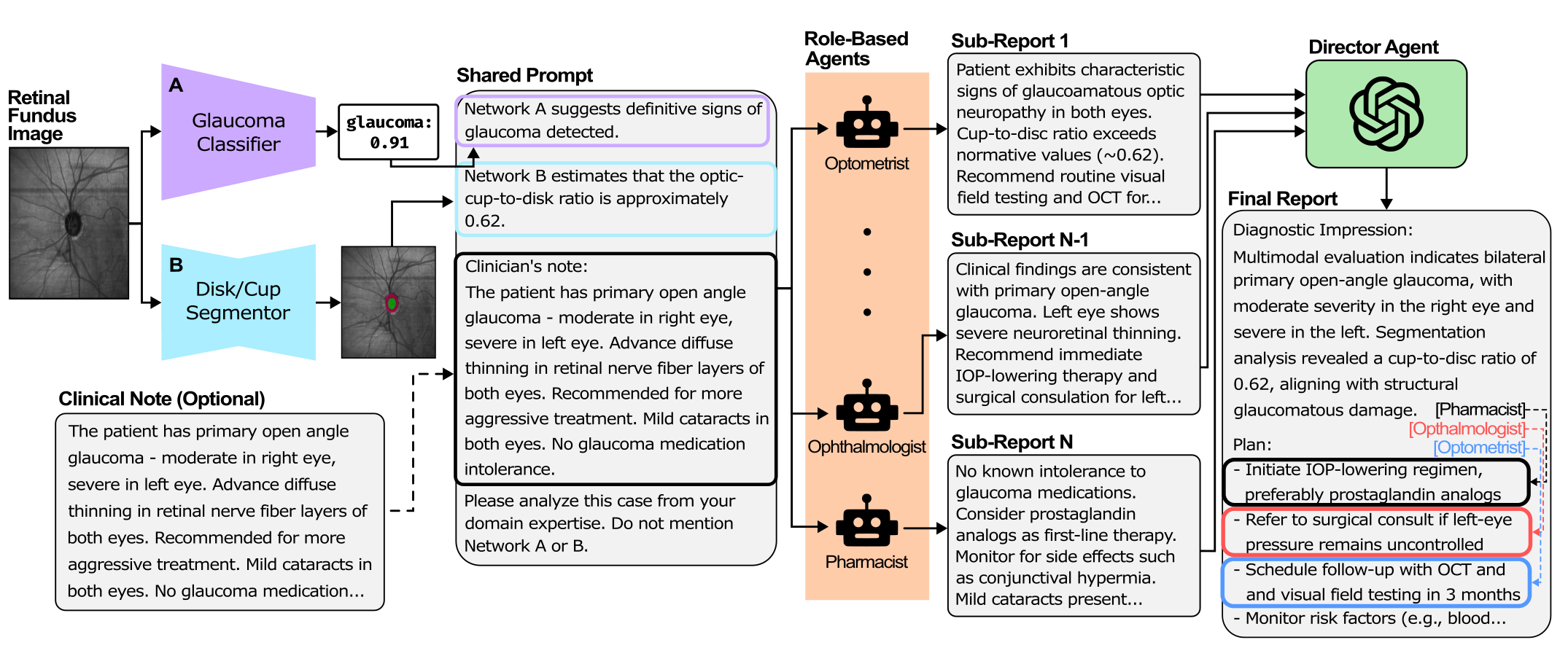}
    \vspace{-6mm}
    \caption{Overview of the MedChat framework. Structured outputs from vision models and clinician notes are processed by role-specific agents and synthesized into a diagnostic report. Fundus image and note from the Harvard-FairVision dataset \cite{luo2024fairvisionequitabledeeplearning}.}
    \vspace{-5mm}
    \label{fig:SL}
\end{figure*}

\subsection{Overview}
We present MedChat, a multi-agent diagnostic system that combines computer-aided diagnosis networks with LLM reasoning to emulate a multidisciplinary clinical workflow. It first extracts structured findings from medical images using deep learning modules, which are then verbalized into natural language. These textual summaries, optionally enriched with clinician-provided notes, are provided to a set of specialized agents, each simulating a distinct glaucoma-relevant clinical role (e.g., ophthalmologist, optometrist). The agents independently generate domain-specific observations, which are synthesized into a cohesive diagnostic report by a director agent. While we demonstrate MedChat for glaucoma diagnosis using retinal fundus images, the framework is broadly adaptable to other imaging-based diagnostic tasks. An overview of the MedChat framework is shown in Fig. \ref{fig:SL}.

\subsection{Input Processing and Prompt Construction}
The MedChat pipeline begins by translating retinal fundus images into clinically structured natural language prompts that are shared across all agents. This prompt is constructed using outputs from two core modules: a classifier estimating the likelihood of glaucoma and a segmentor localizing anatomical structures, specifically the optic disc and optic cup. 

To assess the likelihood of glaucoma, we employ a pre-trained SwinV2-based glaucoma classifier on retinal fundus images~\cite{sun2023swinv2glaucoma, liu2022swintransformerv2scaling}, denoted $C(\cdot)$. Given a fundus image $X$, the model outputs a scalar probability $p = C(X)$ indicating the presence of glaucoma. We then discretize this probability into one of four verbal diagnostic grades using the following mapping:
\[
\text{Grade}(p) =
\begin{cases}
\text{no glaucoma,} & \text{if } p < 0.2 \\
\text{possible glaucoma,} & \text{if } 0.2 \leq p < 0.5 \\
\text{likely glaucoma,} & \text{if } 0.5 \leq p < 0.9 \\
\text{glaucoma detected,} & \text{if } p \geq 0.9
\end{cases}
\tag{1}
\]
Verbalizing the model’s output parallels qualitative phrasing in clinical reports and helps LLMs reason more effectively over diagnostic content \cite{Wang2024}.

Subsequently, we apply a SegFormer-based semantic segmentation model~\cite{segformer_optic_disc_cup, xie2021segformersimpleefficientdesign}, denoted $S(\cdot)$, to localize anatomical structures in a fundus image $X$. The model produces a pixel-wise segmentation map $S(X)$, assigning each pixel a label: optic disc, optic cup, or background. From this map, we extract binary masks for the optic cup and optic disc, denoted $M_{\text{cup}}$ and $M_{\text{disc}}$, respectively.

From these masks, MedChat computes the cup-to-disc ratio (CDR), a widely used quantitative marker for glaucoma diagnosis. Because the optic cup is anatomically contained within the optic disc, we define the total optic disc area as the union of the cup and surrounding disc. Following prior work~\cite{Syc2011-kf}, the CDR is approximated by taking the square root of the ratio between the cup area and the total disc area:
\[
\text{CDR} = \sqrt{\frac{|M_{\text{cup}}|}{|M_{\text{cup}}| + |M_{\text{disc}}|}}
\tag{2}
\]
where $|M_{\text{cup}}|$ and $|M_{\text{disc}}|$ denote the total number of pixels in the optic cup and optic disc masks, respectively. This numerical value is then expressed in natural language, for example: 
\begin{framed}
\textit{“The optic-cup-to-disc ratio is approximately 0.62”}
\end{framed}

To construct the shared core prompt, we label the classifier as Network A and the segmentor as Network B, following the convention introduced in ChatCAD~\cite{Wang2024}. Their natural language outputs are then concatenated to form the prompt. When available, clinical notes, including relevant history, medication use, and exam findings, are appended to the core prompt to enrich the diagnostic context. If no such notes exist, the prompt omits this component. An example core prompt including clinical notes is shown below:

\begin{framed}
\emph{“Network A suggests definitive signs of glaucoma detected. Network B estimates that the optic-cup-to-disc ratio is approximately 0.62. Clinician's notes: \{clinical note\}.”}
\end{framed}

The core prompt, composed of the classifier output, segmentation-derived CDR, and optional clinical notes, forms the factual input for all downstream role-specific agents responsible for generating diagnostic sub-reports.

\subsection{Role-Specific Agent Generation}

In the second stage of the pipeline, MedChat uses the shared prompt to elicit domain-specific insights from multiple role-specific LLM agents. It first queries GPT-4.1 to identify a set of clinically relevant roles appropriate for the diagnostic context defined by the shared prompt. For a given case, roles such as ophthalmologist, optometrist, pharmacist, and glaucoma specialist may be generated. A corresponding GPT-4.1 agent is instantiated for each identified role.

All agents receive the same core prompt, which includes the natural language outputs from the image classifier and segmentor, along with any available clinician notes. This prompt is augmented with role-specific instructions that define the agent’s perspective and constrain its response to information relevant to its clinical expertise. For example, each agent receives a prompt of the following form:

\begin{framed}
\emph{“\{core prompt\}}

\emph{As a \{role\}, please analyze this case from your domain expertise. Only include observations and recommendations relevant to your specialty. Avoid repeating what is not within your scope. Do not mention Network A or B. Write this as part of a professional medical report meant to be integrated with other specialists' insights.”}
\end{framed} 

These instructions are designed to maintain a professional clinical voice, avoid artificial references to model components, and prevent overlap across agent outputs. All agents operate independently, each generating a sub-report aligned with its clinical scope. These sub-reports consist of professional medical observations and recommendations written in the language and tone typical of each role.

By encouraging each agent to contribute findings unique to its specialty, MedChat promotes breadth and minimizes redundancy, allowing the case to be considered from multiple expert perspectives, reflecting the collective expertise of a multidisciplinary medical team.

\subsection{Director-Level Report Generation}

Once the sub-reports are generated, they are aggregated and passed to a final GPT-4.1 instance referred to as the director agent. This agent functions as a medical report synthesizer, tasked with composing a unified diagnostic report that integrates the role-specific findings into a coherent narrative. Its prompt consists of the concatenated sub-reports along with instructions to identify areas of consensus, resolve minor contradictions, and produce a summary in the tone of a formal medical record. 

For example, in a case where a clinical note was provided and four roles were generated, the director agent's prompt is structured as follows:

\begin{framed}
\emph{“The following are diagnostic reports from multiple medical professionals regarding a suspected case of glaucoma based on a fundus image, CAD analysis, and a clinical note.}

\emph{Report: \{sub-report \#1\}}

\emph{Report: \{sub-report \#2\}}

\emph{Report: \{sub-report \#3\}}

\emph{Report: \{sub-report \#4\}}

\emph{Based on the information above, write a final comprehensive diagnostic report. Summarize the key findings, provide diagnostic reasoning, and include appropriate recommendations. Do not reference the sources of the information or mention any sub-reports. The report should be written in a professional, neutral tone suitable for a clinical team or patient record.”}
\end{framed}

Since all agents receive the same core prompt, the director agent is instructed to avoid referencing the sub-reports or their sources explicitly. This ensures the final report remains concise and authoritative in tone, without redundant attributions. It also prevents misleading formulations, such as “all agents agree on a CDR of 0.43,” which may falsely imply that the consensus emerged independently rather than from shared input. By omitting such references, the director-level synthesis appears more direct, cohesive, and clinically appropriate.

The resulting output is a finalized report that presents a structured diagnostic impression, supporting evidence, and actionable recommendations, and is communicated with the clarity and tone expected from a senior clinician. This report serves as the primary output of the MedChat system and is intended for both medical evaluation and patient-facing documentation.

Beyond summarization, the director agent plays a critical role in enhancing system robustness. It can correct minor inaccuracies or inconsistencies that may appear in individual sub-reports and synthesize novel insights across roles into a unified and clinically appropriate plan. This synthesis step not only improves the coherence of the final output but also reflects a key advantage of MedChat's multi-agent architecture: its ability to integrate diverse medical perspectives into a clinically sound and comprehensive diagnostic conclusion.

\subsection{Design Insights}

Through its multi-agent architecture, MedChat emulates the layered reasoning and collaborative synthesis characteristic of multidisciplinary clinical workflows. Domain-specific agents generate focused sub-reports grounded in CAD-derived evidence, which are then aggregated by a coordinating director agent into a comprehensive diagnostic report. By incorporating explicit features, such as glaucoma probability and cup-to-disc ratio, into the core prompt, MedChat ensures that outputs remain anchored in verifiable observations. This design improves medical relevance, accuracy, and transparency, addressing a common limitation of LLM-based diagnostic systems.

MedChat’s modular design also enables flexible adaptation and domain transfer. Vision components, prompt construction logic, and agent configurations can be independently updated, replaced, or expanded without altering the overall architecture. For instance, the segmentation model can be upgraded, LLM agents replaced with fine-tuned variants, or new components, such as a blood vessel segmentor, added to enrich the core prompt. By adjusting modules and agent roles, MedChat can be tailored to different diagnostic objectives.

\begin{figure}[!tp]
    \includegraphics[width=1\linewidth]{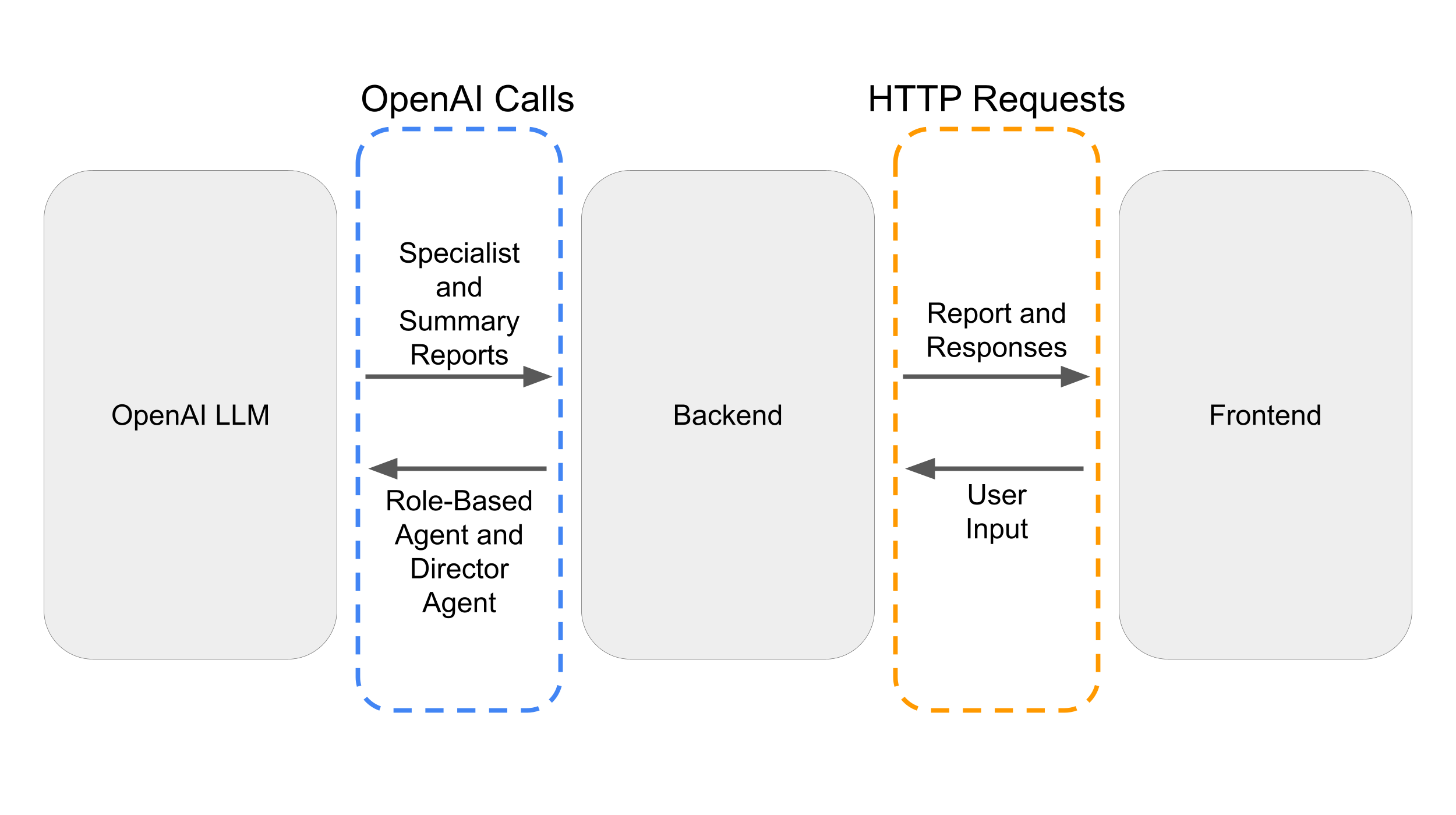}
    \vspace{-10mm}
    \caption{Overview of the MedChat platform architecture. User input is routed through the backend to generate role-based and summary reports via OpenAI API calls, with outputs returned to the frontend.}
    \vspace{-6mm}
    \label{fig:platform-overview}
\end{figure}

\section{Platform Development}

\subsection{Overview}

To demonstrate the functionality of the MedChat framework, we developed a complete end-to-end platform that enables interactive chatbot use and demonstrates system capabilities. This section describes the technical architecture and implementation of the two core components developed in-house: the backend and frontend. Fig.~\ref{fig:platform-overview} illustrates the overall platform design and component interactions.

\subsection{Backend}

The backend of the MedChat platform implements the full diagnostic pipeline in Python, as shown in Fig.~\ref{fig:SL}. It processes retinal fundus images and optional clinical notes through modular stages: deep learning-based image analysis, prompt construction, role-specific agent generation, and director-level report synthesis. Outputs from the classifier and segmentation models are verbalized into natural language and passed to LLM agents aligned with distinct clinical roles. These sub-reports are then synthesized into a comprehensive diagnostic report. The modular design supports extensibility, allowing individual components to be upgraded or replaced without affecting overall functionality or interface compatibility.

In addition to report generation, the backend supports an interactive chat interface for clarifying report content, interpreting clinical terminology, and handling follow-up queries or revised inputs. The chatbot is powered by an instance of GPT-4.1, and conversation history is retained across interactions to preserve context and improve response relevance.

To enable scalable, multi-user deployment, the backend assigns a unique identifier to each session and uses it to maintain isolated interaction histories. All backend functionalities are exposed via RESTful API endpoints to enable seamless integration with the frontend. These endpoints handle incoming user data, route requests through the appropriate pipeline components, including OpenAI calls for report generation or question-and-answer interaction, and return the results for real-time display.

\subsection{Frontend}
\begin{figure}
    \includegraphics[width=1\linewidth]{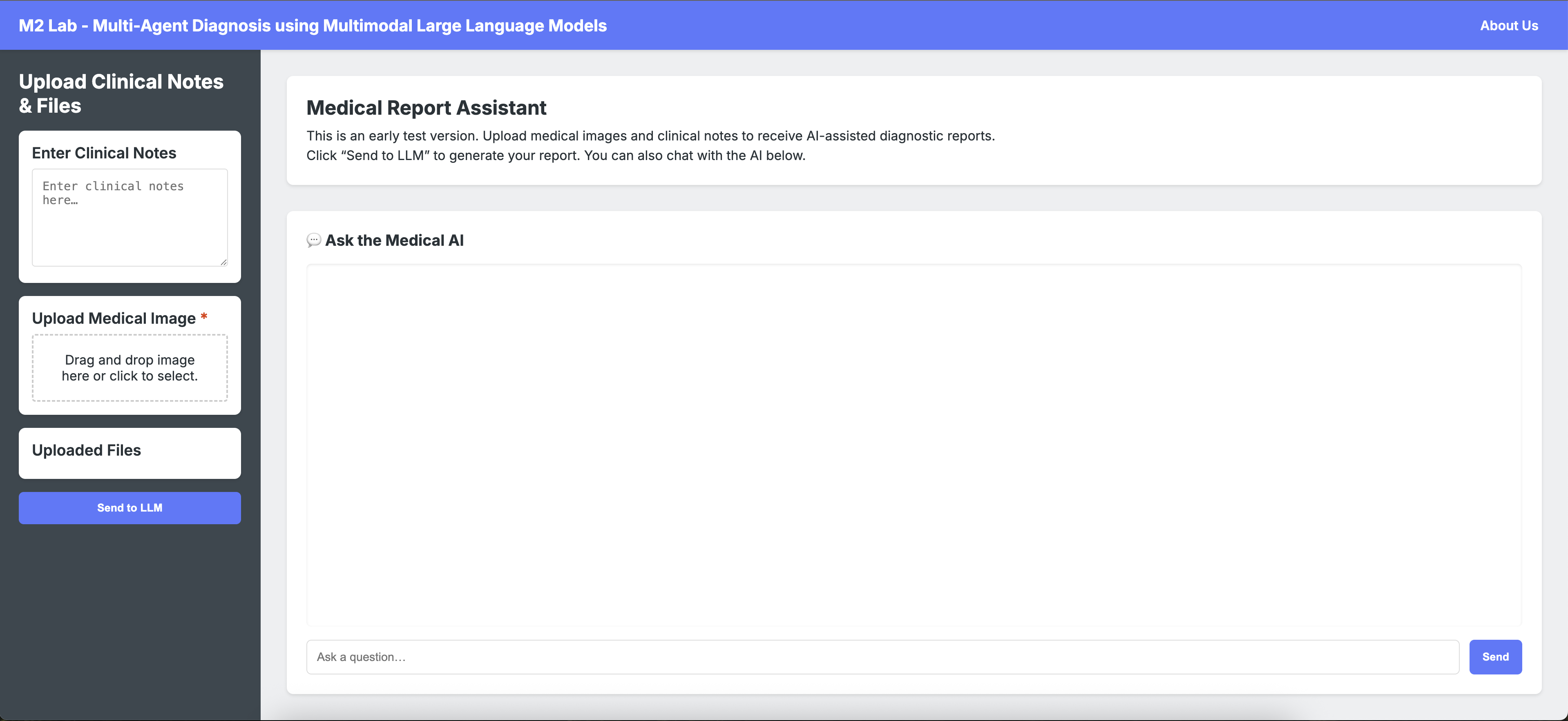}
    \vspace{-5mm}
    \caption{Initial upload interface for MedChat showing input options for fundus image and clinical notes. Users can initiate report generation via the “Send to LLM” button.}
    \vspace{-4mm}
    \label{fig:start}
\end{figure}
\begin{figure}
    \includegraphics[width=1\linewidth]{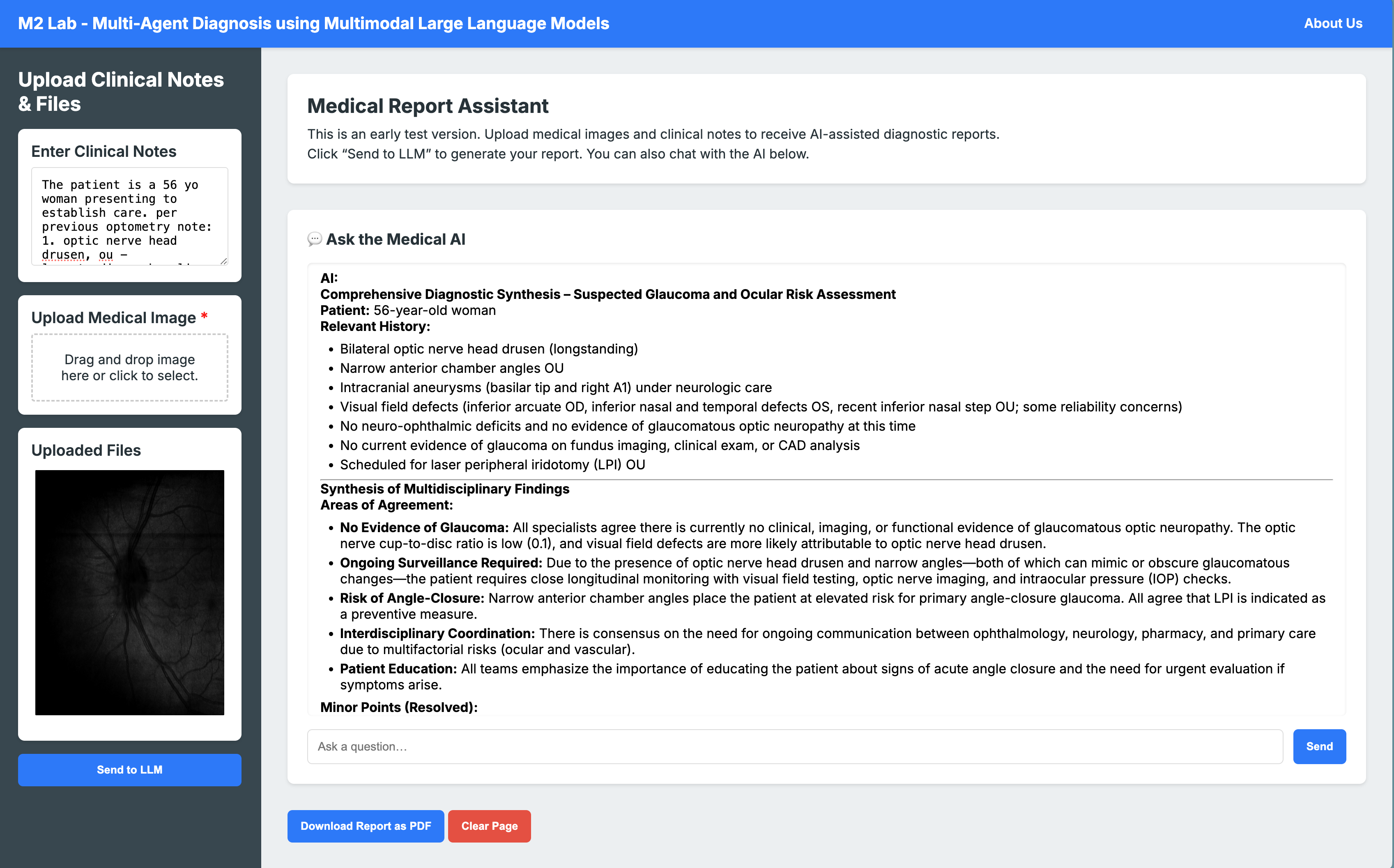}
    \vspace{-5mm}
    \caption{Interface after diagnostic report generation. Users can view the structured summary and export the complete report using the “Download Report as PDF” button.}
    \vspace{-4mm}
    \label{fig:generated}
\end{figure}
\begin{figure}
    \includegraphics[width=1\linewidth]{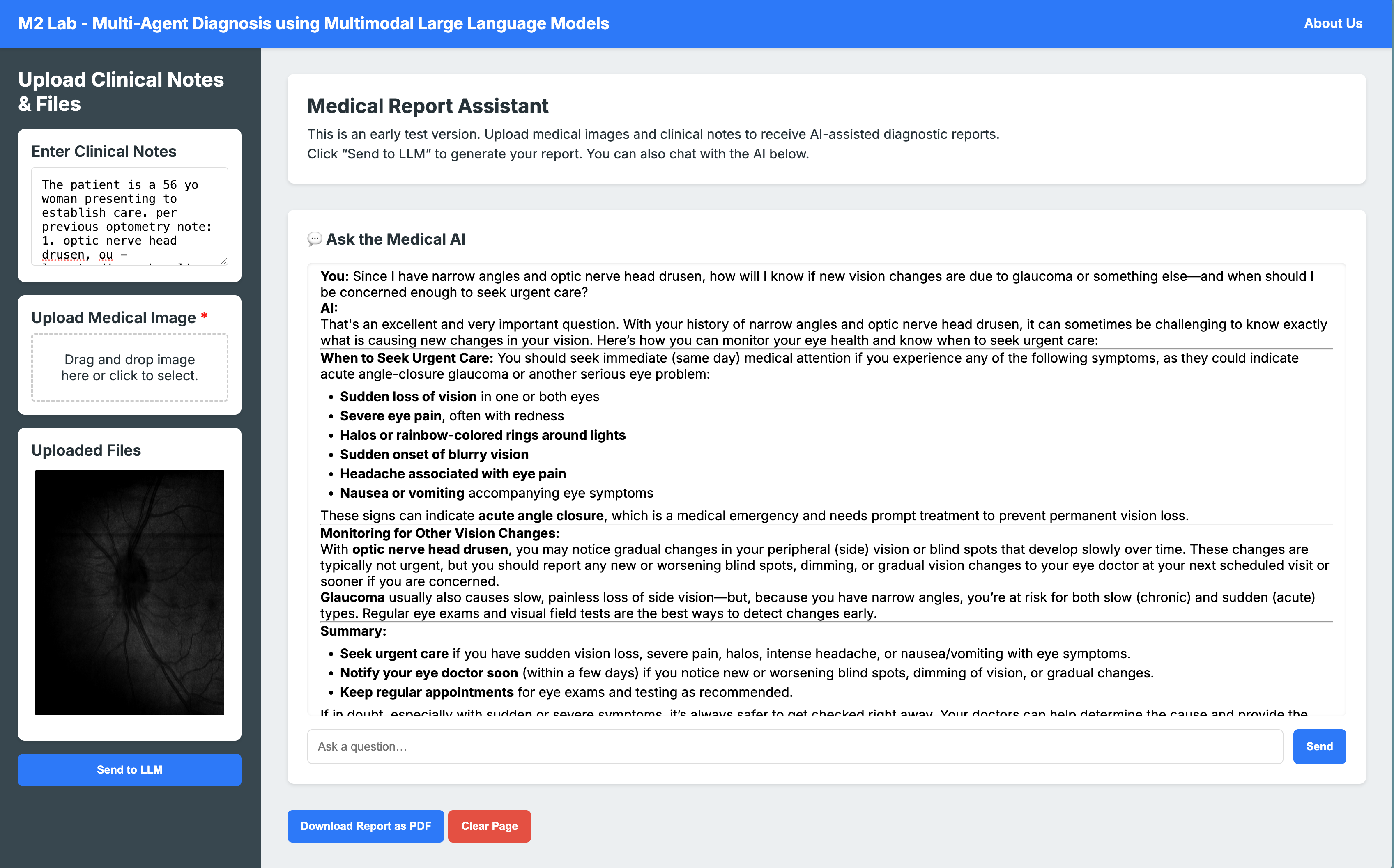}
    \vspace{-5mm}
    \caption{Interface for question-and-answer interaction. Users can ask follow-up questions related to the diagnostic report, and responses are generated in context by the language model.}
    \vspace{-5mm}
    \label{fig:chat}
\end{figure}

The frontend is a lightweight, deployable web application built using JavaScript and HTML. It provides an intuitive interface for uploading retinal fundus images, entering optional clinical notes, generating diagnostic reports, and querying the language model.

Upon page load, users are presented with an upload panel and input field for clinical notes (Fig.~\ref{fig:start}). Submitting this information via the “Send to LLM” button triggers a RESTful API call to the backend, which returns a Markdown-formatted diagnostic report (Fig.~\ref{fig:generated}). A built-in chat interface (Fig.~\ref{fig:chat}) allows users to ask follow-up questions about the report or clinical context.

All results, including the input image with overlaid segmentation masks, the diagnostic report, and the chat transcript, are compiled into a downloadable PDF summary. This design ensures a complete, self-contained record suitable for clinical review or patient communication.

\section{Experiments}
\begin{figure*}[t]
    \centering
    \includegraphics[width=1\linewidth]{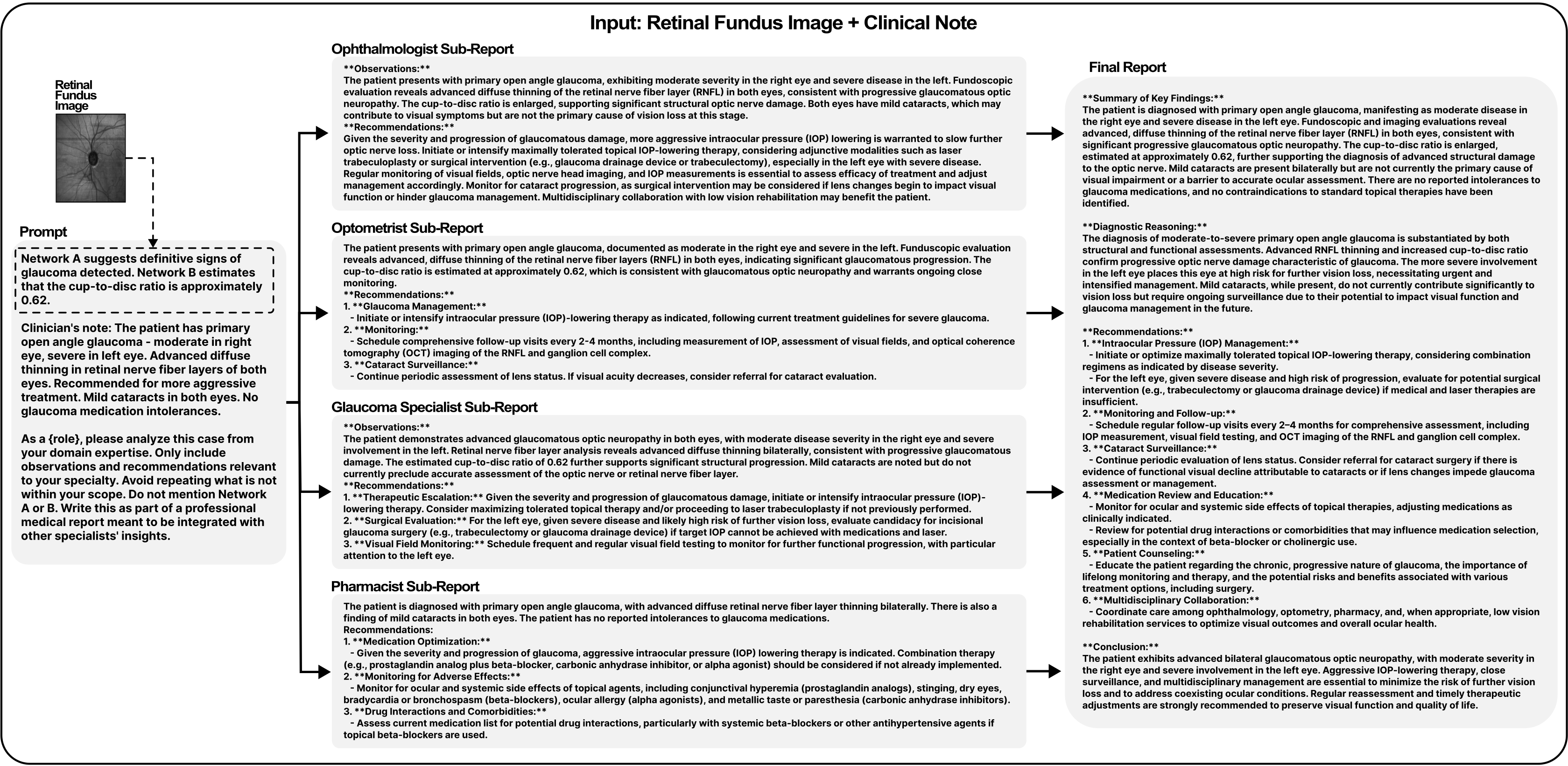}
    \vspace{-6mm}
    \caption{End-to-end diagnostic report generation in MedChat. The figure shows the initial prompt, sub-reports from role-specific agents, and the final synthesized report. Fundus image and clinical note from the Harvard-FairVision dataset \cite{luo2024fairvisionequitabledeeplearning}.}
    \vspace{-6mm}
    \label{fig:7}
\end{figure*}

This section demonstrates how diagnostic text evolves through the MedChat pipeline, using a retinal fundus image and an accompanying clinical note as input.

As illustrated in Figure~\ref{fig:7}, the initial prompt, comprising image-derived CAD outputs, the clinical note, and supplementary instructions, is distributed to multiple agents, each assigned a clinically relevant role. These agents then generate domain-specific sub-reports: the ophthalmologist highlights structural damage and recommends surgical intervention for the left eye; the optometrist stresses the need for follow-up testing and additional imaging; the pharmacist evaluates medication regimens and side effect risks; and the glaucoma specialist provides a comprehensive assessment of disease severity with a multi-step treatment plan.

The director agent then synthesizes these sub-reports into a unified diagnostic report that reflects both image findings and clinical context. It emphasizes the advanced stage of glaucoma in the left eye, recommends escalation of treatment, including possible surgical intervention, and consolidates the follow-up and medication guidance into a clear management plan. The report avoids redundancy, resolves minor inconsistencies, and provides actionable next steps rooted in the severity of the disease. This example illustrates MedChat's core capability to produce diverse clinical perspectives and combine them into a comprehensive diagnostic report. We leave formal evaluation of diagnostic quality, coherence, and clinical accuracy to future work.

\section{Conclusion}

In this paper, we presented a multi-agent diagnostic framework that combines CAD modules with role-specific LLM reasoning to generate rich, clinical reports from retinal fundus images. By distributing diagnostic tasks across specialized agents and synthesizing their outputs through a director agent, the system mirrors medical workflows while improving interpretability and robustness.

Our system has several limitations. First, all LLM agents are used in their general form without domain-specific fine-tuning, which can reduce clinical precision. Second, there is no feedback loop to incorporate clinician oversight—expert corrections made during review are not captured or used to improve the system. Third, since all agents receive the same shared prompt, their outputs often show a high degree of consensus, limiting the diversity of reasoning. Finally, when clinical notes are unavailable, the core prompt lacks sufficient context, leading to more generic responses.

Future work opens up multiple areas to address these challenges. An immediate next step is to fine-tune the LLM agents on disease-specific data, for example, using a corpus of glaucoma case reports and expert-authored guidelines, so that their generated text aligns even more closely with specialist language and clinical expectations. Another promising direction is to introduce a human-in-the-loop refinement loop, where clinicians can review and correct generated outputs. In particular, we are interested in applying reinforcement learning techniques such as Group Relative Policy Optimization, which enables the use of programmable reward functions to optimize agent behavior  \cite{shao2024deepseekmathpushinglimitsmathematical}.

The modular architecture of our system also offers a way to address limitations. Dynamic, role-specific prompt construction can be implemented to increase variance among sub-reports, where each agent receives a tailored subset of the clinical context, curated for its expertise from upstream modules. This would promote productive divergence among role-specific agents. Similarly, in the absence of a clinical note, additional CAD modules can be incorporated to enrich the prompt and provide deeper context.

 Ultimately, MedChat offers a promising step toward more transparent, modular, and collaborative diagnostic AI. By simulating multidisciplinary reasoning and supporting clinical interaction through an extensible platform, our framework lays the groundwork for safer and more interpretable medical report generation in ophthalmology and beyond.

\section*{Acknowledgments} This work is supported by the U.S. National Science Foundation (NSF) under grant IIS-2434967, the National Artificial Intelligence Research Resource (NAIRR) Pilot and TACC Lonestar6, and Purdue John Martinson Honors College’s Breakthrough Research Award.
The views, opinions, and/or findings expressed are those of the author and should not be interpreted as representing the official views or policies of NSF, NAIRR Pilot, and Purdue University.

\bibliographystyle{ieeetr}

\end{document}